\documentclass[12pt]{article}
\begin{document}
\title{Why the astrophysical black hole candidates may not be  black holes at all}
\author{Abhas Mitra\\
Nuclear Research Laboratory, Bhabha Atomic Research Center\\
Mumbai- 400 085, India\\
Email: amitra@apsara.barc.ernet.in, abhasmitra@rediffmail.com
}
\date{}
\maketitle
In a recent paper[1], it has been shown that, there cannot be any rotating
(Kerr) Black Hole (BH) with finite mass in order that  the generic properties
 associated with the symmetries of stationary axisymmetric 
 Einstein equations are obeyed,
 i.e, in order to have $m\ge 0$, we must have $a=0$. 
 In other words, all observed chargeless BHs with finite masses must be
 non-rotating Schwarzschild BHs. Here, by comparing the {\em invariant 4 volume}
 associated with original Kerr metric [2] with the Boyer-Lindquist version
of the same [3], we further find that, stationary axisymmetric {\em vacuum} Einstein
solutions actually correspond to $m=0$ in addition to $a=0$! This means that
if the Kerr solution is a unique one, the Schwarzschild BHs too correspond
to only $m=0$ and therefore the observed BH candidates (BHs) with $m >0$ are not
BHs at all.
This is in agreement some detailed analysis of recent 
observations[4-7] which suggest the the BHCs have strong
intrinsic magnetic moment rather than any Event Horizon.
If one would derive the Boyer-Lindquist metric in a straightforward
manner by using the Backlund transformation, it would follow that
$a= m \sin \phi$, where $\phi$ is the azimuth angle. This relationship
directly confirms the result that for a supposed rotating BH, actually,
both $a=m=0$.

For all 4-D curvilinear coordinate transformations, it is known that $\sqrt{-g} ~d^4 x
= Invariant$, where $g$ is the corresponding metric determinant. By using this basic mathematical tool, it will be seen in the following
that all Kerr Black Holes have the unique mass $m=0$ in addition to the unique rotation
parameter $a=0$[1]. And  since all observed astrophysical objects in particular
the Black Hole Candidates (BHCs) necessarily have finite mass, $m >0$, it would follow
that they may not be BHs at all. And we point out that the BHcs could be Ultra Compact 
Objects with very high surface gravitational redshift $z\gg 1$ 
and touch upon several  generic properties of
them.

We first recall the original Kerr metric [2]
\begin{eqnarray}
ds^2  &= & \bar\rho^2(d\bar\theta^2 + \sin^2\bar\theta d\bar\phi^2) +
2(d\bar r +d\bar t+a \sin^2\bar\theta d\bar\phi)(d\bar r + a\sin^2\bar\theta d\bar\phi) \nonumber\\
& & -(1-2m\bar r/\bar\rho^2)
(d\bar r + d\bar t+a \sin^2\bar\theta d\bar\phi)^2 ,
\end{eqnarray}
where $\bar t$ is time, $\bar r$ is radial coordinate, $\bar \phi$ is azimuth angle,
$\bar\theta$ is polar angle, $a$ is an integration constant, interpreted as angular momentum per
unit mass; and $m$ too is an integration constant.  And  the parameter
\begin{equation}
\bar\rho^2 = \bar r^2 + a^2 \cos^2{\bar\theta}
\end{equation}
It is implicitly {\em assumed} that the
integration constants $a$ and $m$ 
are positive. 

By means of a straightforward but lengthy algebra, it can be seen the determinant
associated with this metric is (see Appendix I)
\begin{equation}
\bar g = - \bar\rho^4 \sin^2\bar\theta
\end{equation}

By using the following coordinate transformations
\begin{equation}
dt = d\bar t -{2 m r\over \bar\Delta} d\bar r,
\end{equation}

\begin{equation}
d\phi = d\bar\phi + {a\over {\bar\Delta}} d\bar r
\end{equation}
and
\begin{equation}
\theta = \bar\theta; \qquad r =\bar r,~ \rho^2 =r^2 + a^2 \cos^2\theta =\bar r^2 + a^2 \cos\bar\theta^2 =\bar\rho^2
\end{equation}

\begin{equation}
\bar\Delta= \bar r^2 - 2m\bar r +a^2 = r^2 - 2mr +a^2 =\Delta
\end{equation}
Boyer and Lindquist[3] rewrote metric (1) as
\begin{eqnarray}
ds^2 & = & {\rho^2\over \Delta} dr^2 + \rho^2 d\theta^2 +[r^2 +a^2 +{ 2mr\over \rho^2} a^2 \sin^2 \theta]
\sin^2\theta d\phi^2 \nonumber\\
& & + {4amr\over \rho^2} \sin^2\theta d\phi dt -(1- {2mr\over \rho^2}) dt^2
\end{eqnarray}
The Boyer- Lindquist form of the Kerr metric is now considered as
the ``standard'' form for Kerr BHs and in this case, it is already well known[8] that the associated
determinant is
\begin{equation}
g = -\rho^4 \sin^2 \theta = - \bar\rho^4 \sin^2\bar\theta = \bar g
\end{equation}
 Since for {\em all} stationary axisymmetric solutions of Einstein
equations in the ``standard'' form, $g_{\phi \phi} = \sin^2\theta 
g_{\theta \theta}$ when $\theta$ is uniquely defined and
measured from the axis of symmetry, it was found in Paper[1] that either

 \begin{equation}
a=0;\qquad  m \ge 0
\end{equation}
or,
\begin{equation}
a\ge 0; \qquad 2mr =-\rho^2; \qquad m\le 0, \qquad if ~ r\ge 0
\end{equation}
 The latter solution involving negative mass can be seen to be unphysical
in the following manner:

For a given value of $m$ and $a$, Eqs. (6) and (11) would yield

\begin{equation}
r = - m \pm \sqrt{m^2 - a^2 \cos^2\theta}
\end{equation}
This is the equation of a single surface and the 4-D spacetime, thus,
 would collapse to a
3-D spacetime if Eqs.(11-12) would be  valid solutions. If so, then it would be possible to
choose a coordinate system in which $dx^1 = 0$ 
and to set the invariant 4-volume [8,9]
$\sqrt{- g} ~d^4 x =0$.
But since for a 4-D problem, 
 $\sqrt{- g}~ d^4 x >0$ and {\em invariant} too, we must reject
this branch corresponding to negative mass. In any case, we are interested in the
observed BH Candidates (BHCs)   having $m >0$ and hence this $m \le 0$ solution is irrelevant here.

When $a=0$, the transformation equation (5) trivially becomes
\begin{equation}
d \phi = d\bar\phi
\end{equation}
Now if we again use the principle of invariance of 4-volume[8,9] to the coordinate systems
$\bar t, \bar r, \bar \theta, \bar \phi$ and $t, r, \theta, \phi$, we will have
\begin{equation}
\sqrt{- \bar g}~ d\bar t ~d\bar r ~d\bar \theta ~d\bar \phi = \sqrt{-g} ~dt~ dr ~d\theta~d\phi
\end{equation}

But since, now, $d \bar \phi= d\phi$ and already $\bar r = r$, $\bar \theta =\theta$, $\bar g = g$,
we also have

\begin{equation}
d\bar t = d t
\end{equation}
Then  using Eq.(15) in Eq.(4), it trivially follows that,
 in order that stationary axisymmetric {\em vacuum}
Einstein equations obey all the symmetry constraints (Paper I) 
and allowed  transformations, the {\em
integration constant} 
\begin{equation}
m=0
\end{equation}

 in addition to $a=0$. The fact that $a=m=0$ for all Kerr BHs actually {\em follows from
a previous work}:

Neugebauer[10] obtained the Kerr solution from a ``seed solution'' 
by using Backlund 
transformations; the 4th line from the bottom of  p.(73)
of Ref.[10] clearly shows that $a$ and $m$ are related through
\begin{equation}
a = m \sin \phi
\end{equation}
Since $\sin \phi$ is a variable and {\em not identically zero}, the foregoing relation can be satisfied
only when $a=m=0$! However, it is astonishing to see that  
 Neugebauer failed to realize this. Neugebauer's Eq.(17) 
also confirms both the present result ($m=0$) as well as the result of
 Paper[1]
that $a=0$ for Kerr BHs. (Scanned image (pdf) of this page of Ref[10]
is given as Supporting online material II)
Therefore if the Schwarzschild BHs
 are seen to be
arising from a solution of {\em algebrically special}  stationary axisymmetric
vacuum solutions of Einstein equations, then mass of all Schwarzschild BHs would be zero.
However if it would be argued that spherically symmetric vacuum solution exists independent of
general stationary axisymmetric solutions, it would appear that, despite the present exact result,
there could be finite mass Schwarzschild BHs. But this would be highly unlikely
because a spherically symmetric vacuum solution indeed follows as a 
subset of the rotating solution with $a=0$. Furthermore,
 the Kerr solution is considered to be the unque solution.

Therefore the observed BHCs, rotating or non-roating, may not be even Schwarzschild BHs.
Thus the observed BHCs cannot be BHs at all and they cannot have any Event
Horizon (EH), the exclusive hall mark of a BH. 
Whenever there is no EH,
 any astrophysical plasma, even when it is macroscopically neutral,
is likely to develop strong intrinsic magnetic moment upon compactification {\it a la}
 the pulsars  following approximate magnetic flux conservation.
 Thus as per our analysis, the stellar mass BHCs may have locally measured
intrinsic surface magnetic field $B_{local}$  which could be even much stronger than the typical
pulsar magnetic fields. If a compact object has a gravitational surface redshift $z$,
it can be shown that a distant observer would  perceive a surface magnetic field[4-7]
\begin{equation}
B= B(\infty) = (1+z)^{-1} B_{local}
\end{equation}
However, in case, $z \gg 1$, the value of $B$ for such Ultra Compact Objects (UCOs) could
appear weaker than typical pulsar values (which have $z\sim 0.1 -0.2$). 
We recall here that,  for isotropic self-gravitating objects in {\em strict} hydrostatic
equilibrium $z <2$[9,11]. But the BHCs may not be in {\em strict} hydrostatic
equilibrium; they may be  in quasistatic equilibrium at best and  slowly contracting
like primordial clouds and pre-main sequence stars. If so, there need not be any upper limit
on $z$. It may be also recalled here that primordial clouds or pre-main sequence stars
may be in unstable quasiequilibrium (though actually collapsing) for hundreds of million years
without any nuclear burning at their cores. Whenever they are not in {\em strict} equilibrium,
they may be contracting at unimaginably slow rate and by virtue of virial theorem,
while part of the gravitational energy released by contraction goes into generating internal
energy (and attendant pressure gradient), part of it is radiated out[9,11].

One may wonder here ``what about the upper mass limit of $\sim 3 m_\odot$ of compact objects?
This upper mass limit corresponds to {\bf cold} baryonic objects in {\bf strict} hydrostatic equilibrium.
Even when the object is baryonic but {\bf not cold}, this upper limit has little relevance.
As an extreme case of {\bf hot} baryonic objects, one may recall the theoretical possibility
of existence of Supermassive Stars whose masses could be as large as $10^{10} m_\odot$ or even
higher[9,11]. When such ``hot'' objects are not in strict equilibrium, they can generate
their own pressure gradient by virtue of virial theorem.
Also even for an almost cold self-gravitating object like a primordial gas cloud, there
is no upper mass limit when it is not in strict hydrostatic equilibrium.

We may think of
some generic properties of UCOs with $z\gg 1$:

It is known that compact objects with local surface magnetic field considerably higher that $10^{9-10}$ G,
do not display Type I X-ray burst even when they possess a physical surface and this is the reason
that Her X-1 or many other X-ray pulsars do not show Type I X-ray burst. Thus high $z$
UCOs also may not show any Type I burst activity despite possessing a physical surface.

When $z \gg 1$, any signal generated on the surface would propage out through a gravitational
field with extremely steep spatial gradient. Then the temporal properties of the original
signal would be constantly distorted; for example a proper time interval of $\sim 1$s
 on the surface
may appear as $\sim 10$s at few Schwarzschild radii away in case $z \sim 10$. Thus {\em no spin pulsation} would be
directly seen!

On the other hand, strong local intrinsic magnetic field would focus both inflows and outflows. This focusing of
outflow coupled with rotation is likely to generate strong collimated jets; thus ``jets'' could be a generic feature
of such UCOs. As to the generation of Ultra Relativistic jets, 
so far the popular idea has been that the speed of the jet may be bounded by the ``escape velocity'' from the
central object[12]. Thus it was generally believed that jets associated with neutron stars
($1+z\sim 1.1 -1.2$) may not have bulk Lorentz facor higher than $ \Gamma \sim 1.3$ or so whereas jets associated with BHs
($1+z = \infty$) could have $\Gamma \gg 1$. However as of now, observationally, in a strict sense, no jet or no outflow
can be directly associated with any EH or any ``Ergosphere'' because no ``EH'' has been 
detected so far[13]. On the other hand, contrary to the popular idea, it is now known
that Cir X-1, an object {\em with a phyical surface} and strong {\em intrinsic} magnetic field
does give rise to ultrarelativistic outflow with $\Gamma \sim 10$[12]. Thus it should
not be
surprising if it is found that microquasars or quasars with strong outflows contain
central objects having physical surface and intrinsic magnetic field because we have found that objects with EHs have
the unique property $a=m=0$. In fact what could be surprising is the notion that
ultrarelativistic or any oultflow at all could originate from objects {\em 
from which even light cannot escape}.

There could be yet another generic property of an object with high $z$. It is known that
the last stable circulat orbit around a compact object lies at $r= 6 m$ for material particles
and at $r=3m$[11] for photons or other massless particles. The latter value corresponds to
$z \sim 0.72$ which means that once a collapsing object attains $z > 0.72$, photons and neutrinos would find it
extremely difficult to diffuse out from its core. In other words such collapsing would
be extremely ``hot''  and would tend to remain so for long durations.

In fact the detail analysis of X-ray and radio data of the 
stellar mass BHCs indeed suggest that these objects have strong intrinsic magnetic moment and they are
in {\em quasistatic} state probably because of strong magnetic and radiation pressure[4-7]. Robertson \& Leiter
have inferred that the intrinsic luminosity of these objects are close to corresponding
Eddington values yet some of them appear so faint because in GR, the Eddinton luminosity
seen by a distant observer[11] is 
\begin{equation}
L_{ed} (\infty) = 1.3~ 10^{38} ~(1+z)^{-1} ~\left ({m\over 1 m_\odot}\right)~erg/s
\end{equation}
and these objects might have $z \sim 10^{7-8}$. Such a range of value of $z$ may appear outlandish but
here it should be remembered that {\em these range of $z$ or any finite value of $z$ are
 infinitely smaller than the corresponding value of $z=\infty$ for a BH!}
Another explanation for the extremely weak quiesent luminosities of several BHCs and Neutron Stars as well as
the difference in the luminosities between the two categories could be that
the quiesent states are dominated by jet outflows rather than any advection dominated
inflow[14]. Irrespective of such tentative explanations and uncertainties about the
precise nature of BHCs, the fact remains that they have $m >0$
 and hence they maynot be BHs which are characterized by $a=m=0$. 
As a final comment, the result mentioned here, has been obtained independently
in Paper I and the present paper by starting from the equation[15,16]. 
\begin{equation}
g_{\phi \phi} = g_{\theta \theta} \sin^2\theta
\end{equation}
As discussed in Paper I, this relationship is valid for any stationary axisymmetric
metric which is in {\em standard form}, i.e., the only cross term in the metric is
$d\phi dt$ and when $\theta$ is uniquely defined by measuring it from the axis of symmetry
 The html file
for Ref 16 showing this result as well as the pdf file showing the Eq.$a = m \sin \phi$
can be obtained either from the editor or the author.


\section{Appendix 1. Determinant  for the original Kerr Metric}

In its original form, the Kerr metric is[1]
\begin{eqnarray}
ds^2  &= & \rho^2(d\theta^2 + \sin^2\theta d\phi^2) +
2(dr +d t+a \sin^2\theta d\phi)(d r + a\sin^2\theta d\phi) \nonumber\\
& & -(1-2m r/\rho^2)
(d r + d t+a \sin^2\theta d\phi)^2 
\end{eqnarray}
In the parent manuscript, we indicated the original Kerr coordinates by an
overbar which we have removed here.
Here  the parameter
\begin{equation}
\rho^2 =  r^2 + a^2 \cos^2{\theta}
\end{equation}
Let us call
\begin{equation}
x = 1 + {2 mr\over \rho^2}
\end{equation}
Then it can be seen that the various non-vanishing  components of the metric tensor are

\begin{equation}
g_{\theta \theta} = \rho^2
\end{equation}
\begin{equation}
g_{\phi \phi} = (\rho^2 + x a^2 \sin^2 \theta) \sin^2\theta
\end{equation}
\begin{equation}
g_{rr} = x
\end{equation}
\begin{equation}
g_{tt} = -(1- 2mr/\rho^2) =x-2
\end{equation}
\begin{equation}
g_{r \phi} = g_{\phi r} = a x \sin^2 \theta
\end{equation}
\begin{equation}
g_{\phi t} = g_{t \phi} = { 2amr \sin^2 \theta\over \rho^2} = a (x-1) \sin^2\theta
\end{equation}
\begin{equation}
g_{rt} = g_{tr} = {2mr \over \rho^2} = x-1
\end{equation}

The determinant of the metric tensor can be found to be

\begin{eqnarray}
g & = & g_{\theta \theta}[ g_{tt} (g_{rr} g_{\phi \phi} - g_{\phi \phi}^2)
- g_{tr}(g_{tr} g_{\phi \phi} - g_{\phi r} g_{t \phi}) \nonumber\\
& & + g_{t \phi} (g_{rr} g_{t \phi} - g_{tr} g_{r \phi})]
\end{eqnarray}
First note that

\begin{equation}
(g_{rr} g_{t \phi} - g_{tr} g_{r \phi})= a(x-1) x\sin^2\theta - (x-1)
 a \sin^2\theta =0
 \end{equation}
 so that
 \begin{equation}
g  =  g_{\theta \theta}[ g_{tt} (g_{rr} g_{\phi \phi} - g_{\phi \phi}^2)
- g_{tr}(g_{tr} g_{\phi \phi} - g_{\phi r} g_{t \phi})] 
\end{equation}
Then using Eqs.(22-29) in (32), we have,
\begin{eqnarray}
g/\rho^2 & = & (x-2)[ x \rho^2 \sin^2\theta + x^2a^2 \sin^2\theta- a^2 x^2 \sin^\theta]\nonumber\\  
     & & - (x-1) [(x-1)  \sin^2\theta (\rho^2 + x a^2 \sin^2\theta)-a^2x(x-1) \sin^4\theta] \nonumber\\
& = & (x-2) x \rho^2 \sin^2\theta\nonumber\\
& & -(x-1)^2 \sin^2\theta [\rho^2 + xa^2 \sin^2\theta - xa^2\sin^2\theta]\nonumber\\
& = & (x-2) x \rho^2 \sin^2\theta - (x-1)^2 \sin^2\theta \rho^2\nonumber\\
& = & \rho^2 \sin^2\theta [x(x-2) - (x-1)^2] \nonumber\\
& = & -\rho^2 \sin^2\theta
\end{eqnarray}

Therefore the
determinant 
\begin{equation}
g = -\rho^4 \sin^2 \theta 
\end{equation}
Now introducing overbar for the original Kerr metric, as has been done in the parent manuscript, we obtain
\begin{equation}
\bar g = - \bar\rho^4 \sin^2\bar\theta
\end{equation}

\end{document}